\documentclass[journal]{IEEEtran}

\usepackage{blindtext}
\usepackage{graphicx}
\usepackage{pbox}
\usepackage{multirow}
\usepackage{algorithm,algorithmic}
\usepackage{svg}
\usepackage{booktabs}

\usepackage{url}
\usepackage{listings}
\usepackage{color}
\definecolor{lightgray}{rgb}{.9,.9,.9}
\definecolor{darkgray}{rgb}{.4,.4,.4}
\definecolor{purple}{rgb}{0.65, 0.12, 0.82}
\newcommand{\LineComment}[1]{ \hfill/* \textit{#1} */}
\newcommand{\LineCommentLeftAlign}[1]{/* \textit{#1} */}

\lstdefinelanguage{JavaScript}{
  keywords={typeof, new, true, false, catch, function, return, null, catch, switch, var, if, in, while, do, else, case, break},
  keywordstyle=\color{blue}\bfseries,
  ndkeywords={class, export, boolean, throw, implements, import, this},
  ndkeywordstyle=\color{darkgray}\bfseries,
  identifierstyle=\color{black},
  sensitive=false,
  comment=[l]{//},
  morecomment=[s]{/*}{*/},
  commentstyle=\color{purple}\ttfamily,
  stringstyle=\color{red}\ttfamily,
  morestring=[b]',
  morestring=[b]"
}

\begin{document}
\title{An Unsupervised Approach to Detect Spam Campaigns that Use Botnets on Twitter}

\author{Zhouhan~Chen,~\IEEEmembership{Rice University,}
        Devika~Subramanian,~\IEEEmembership{Rice University}% <-this % stops a space
}

%\markboth{Journal of \LaTeX\ Class Files,~Vol.~6, No.~1, January~2007}%
%{Shell \MakeLowercase{\textit{et al.}}: Bare Demo of IEEEtran.cls for Journals}

\maketitle

\section{Abstract}

In recent years, Twitter has seen a proliferation of automated accounts or bots that send spam, offer clickbait, compromise security using malware, and attempt to skew public opinion \cite{Thomas:SuspendedAccountsinRetrospect, FirstMonday}. Previous research estimates that around 9\% to 17\% of Twitter accounts are bots contributing to between 16\% to 56\% of tweets \cite{Varol:OnlineHumanBotInteractions, Chen:HuntingMaliciousBots} on the medium. This paper introduces an unsupervised  approach to detect Twitter spam campaigns in real-time. The bot groups we detect tweet duplicate content with shortened embedded URLs over extended periods of time. Our experiments with the detection protocol  reveal that bots consistently account for 10\% to 50\% of tweets generated from 7 popular URL shortening services on Twitter. More importantly, we discover that  bots using shortened URLs  are connected to large scale spam campaigns that control thousands of domains. There appear to be two distinct mechanisms used to control bot groups and we investigate both in this paper. Our detection system runs 24/7 and actively collects bots involved in spam campaigns and adds them to an evolving database of malicious bots. We make our database of detected bots available for query through a REST API so others can filter  tweets from malicious bots to get high quality Twitter datasets for analysis.

%We discover several spam campaigns that control over thousands of domains and use Twitter botnets to spread malicious but shortened links. We also discover that bot makers have evolved in recent years, from controlling a large number of bot accounts to building a single Twitter application to hijack normal users' credentials. Those two mechanisms are closely studied in this paper with evidence collected by our system. 

%This detection system is able to farm (a) bot groups tweeting shortened URLs by running scheduled crawling jobs and (b) bot groups tweeting non-shortened URLs by monitoring suspicious trending URLs on Twitter. After identifying botnets, we take one step further to resolve final landing URLs behind botnets to identify social spam campaigns.

%Our month-long experimental results show that bots consistently account for 10 to 50\% of tweets generated from 7 popular URL shortening services on Twitter. 

%Our detection system runs 24/7 to collect new botnets and spam campaigns and share the intelligence to researchers to clean Twitter datasets and improve Internet security.

% \fi
\IEEEpeerreviewmaketitle

\section{Introduction}
Twitter bots are accounts operated by programs instead of humans \cite{Ferrara:TheRiseofSocialBots}. Previous research estimates that 9\% to 17\% of Twitter accounts are bots, and that between 16\% to 56\% of tweet volume on Twitter is generated and propagated by the bot population. %Not all bot accounts are created equal. 
Although some bot accounts such as Earthquake Bot\footnote{Earthquake Bot: \url{https://twitter.com/earthquakebot?lang=en}} are beneficial to the community, many are malicious. Our work focuses on malicious bot groups that send tweets containing embedded URLs.  URL-tweeting bots are widely used by promotional campaigns to gain traffic to their websites, by spammers to redirect users to phishing sites and by malicious users as click bait to spread malware \cite{Zhang:DetectingSpam}. Prior work  analyzing large Twitter datasets concludes that most bots include URLs in tweets and that the average number of URLs in tweets sent by bots is three times higher than the number in tweets from human accounts \cite{Chu:HumanBotCyborg}. Accounts tweeting  embedded URLs over extended time periods are more likely to be  malicious and 
%be  suspended by Twitter for 
are more likely to violate Twitter's terms of service\footnote{The Twitter Rules: \url{https://support.twitter.com/articles/18311}}.
% than accounts tweeting plain text. 
Systematically detecting such malicious accounts and flagging them for suspension or elimination is the goal of our work.

Existing work on detecting Twitter spam campaigns is either retrospective \cite{Thomas:SuspendedAccountsinRetrospect}, where the analysis is done after accounts are suspended by Twitter, or prospective, in which detection is done while the accounts are still active. Most prospective detection methods are supervised \cite{Chu2012:DetectingSocialSpamCampaigns} or semi-supervised \cite{Sedhai:SemiSupervisedSpamDetection}. Even though some of these approaches  focus on identifying collective action among multiple accounts through the use of embedded URLs, they still require manual labeling of accounts to train classifiers for detecting malicious bots. Our system, on the other hand, actively farms new spam campaigns without any human intervention. Our completely unsupervised detection approach enables us to build a  blacklist of malicious email addresses, URLs and Twitter accounts, and to share threat intelligence with the research community in real-time.

%Once our system detects botnets, it will proceed to identify spam campaigns by extracting suspicious registrants who register multiple domains and use botnets to send URLs that redirect to those domains

Section \ref{section:method} introduces our unsupervised, real-time Twitter botnet and spam campaign detection system. Section \ref{section:Experimental_Results} shows results of our month-long experiment and compares our approach with existing bot detection methods. Section \ref{section:bot_mechanism} details two mechanisms that spammers use to control Twitter bot groups. Section \ref{section:conclusion} summarizes our contribution.

\section{Method} \label{section:method}
\begin{figure*}[!ht]
\centerline{\includegraphics[width = 0.97 \textwidth]{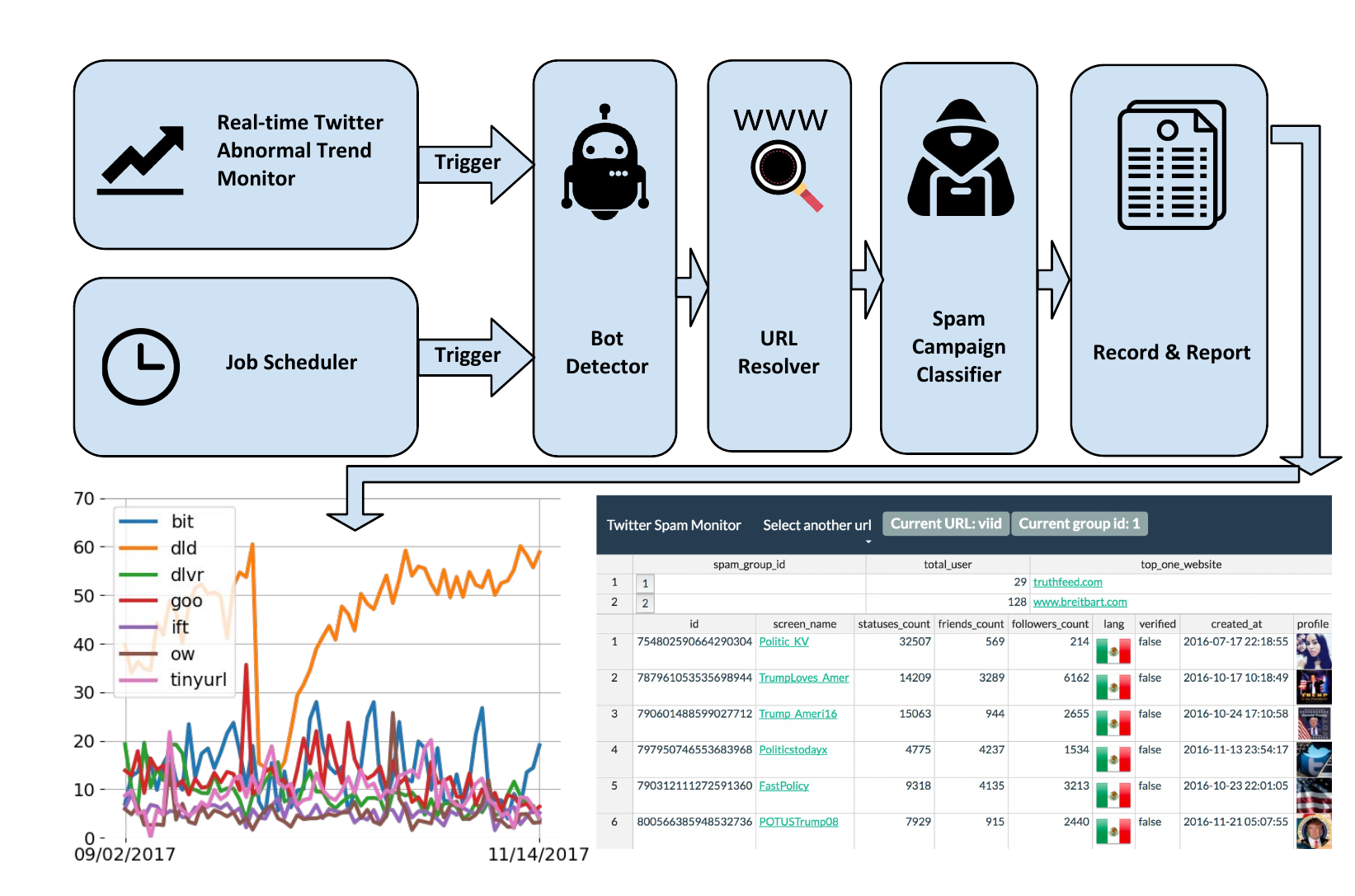}}
\caption{Top: Real-time Twitter Bot Detection System.
Bottom left: Percentages of daily tweets created by bots. The most abused URL shortening service is {\em dld.bz}. One of the botnets using {\em dld.bz} stopped tweeting for several days in the middle of September, which accounts for the sudden dip in the {\em dld.bz} curve.
Bottom right: Our Twitter Bot Monitor  shows a group of accounts that superficially resemble  Trump supporters.Their real objective is to inject malware by tricking users to click on shortened URLs. The link to our Twitter Bot Monitor is \protect\url{http://water.clear.rice.edu:18000}}
\label{figure:architecture}
\end{figure*}

The bot detection system has a backend subsystem which is responsible for detection and a frontend subsystem that serves up results through application program interfaces (APIs) and through visualizations. The backend subsystem consists of five parts: a real-time Twitter abnormal trend monitor, a job scheduler, a bot detection routine, a URL resolver and a spam campaign classifier. The URL monitor collects and monitors the top $k$ URLs found in  tweets on Twitter's real-time Streaming API,
%unshortened suspicious URLs 
and the job scheduler collects shortened URLs. Both types of tweet collections are passed through a pipeline consisting of the bot detection routine, the URL resolver and the spam campaign classifier.

The frontend consists of a web application and a set of APIs. The web interface visualizes bot profiles and key statistics in a table format and the API sends bot statistics in JSON format. Both services are available to researchers who want to access archived bot accounts found by our system.

\subsection{Backend Subsystem}
%use of netloc ensures that all generalizability \cite{python:urllib}}.
\subsubsection{Real-time Twitter abnormal trend monitor}
The real-time Twitter trend monitor runs 24/7 to detect abnormal signals in the Twitter stream. Currently, we use the keyword {\em http} to collect any tweet that contains a URL. A counter in our  system stores and updates occurrences of the network location part\footnote{A general URL has a form \url{scheme://netloc/path;parameters?query#fragment}} (netloc) of each URL collected. Using netloc gives us the ability to perform a finer-grained  subdomain-level analysis of trending URLs. This is particularly useful because as we will show in Section \ref{section:method}, spammers abuse subdomain registration to create spam campaigns. Our system keeps a white-list of popular and trustworthy URLs, and anytime a non-white-listed URL enters the  top $k$ list, our system automatically triggers the bot detection routine and uses the network location part as the input keyword to gather more tweets. The top $k$ counter is reset after $n$ minutes to ensure that recent trending URLs have a chance to enter the top $k$ list. In our deployed system we use $k=15$ and $n=60$. A larger $k$ and a smaller $n$ will capture more bot activities, but the scale of the botnets (number of accounts in a botnet, number of tweets generated by a botnet) detected will be smaller.

\subsubsection{Job scheduler}
The job scheduler schedules and triggers a bot detection routine based on a predetermined list of keywords at a predefined time interval. Currently, our system collects bot groups from the seven most popular URL shortening services on Twitter ({\em bit.ly, goo.gl, ow.ly, tinyurl.com, dlvr.it, if.ttt, dld.bz}) every 24 hours. If a new URL shortening service is detected by our abnormal trend monitor, we  update the list of  URLs to be tracked by the job scheduler, which will automatically begin the botnet detection protocol on  the newly added URL.

\subsubsection{Bot detector}
The bot detector consists of four parts. A \textbf{crawler} collects live tweets based on input keywords. A \textbf{duplicate filter} hashes each tweet text to a group of users and identifies groups that tweet the same text. Groups with 20 or more accounts are passed to the next stage. The threshold 20 ensures that only important groups are considered. A \textbf{collector} then calls Twitter Rest API to collect the most recent 200 tweets from every account in a group, and calculates an overlap ratio for each account between that account's tweets and other tweets. If the ratio is higher than a threshold, the account is flagged as a bot. This step ensures that we filter out innocent users who happen to tweet the same text as bots. A {\bf detector} uses an unsupervised clustering algorithm to find groups of accounts tweeting texts with high similarity. Algorithm~\ref{alg:bot_detection} shows the detailed clustering steps.

 \begin{algorithm}
 \caption{Algorithm for detecting botnets}
 \label{alg:bot_detection}
 \begin{algorithmic}[1]
 \renewcommand{\algorithmicrequire}{\textbf{Input:}}
 \REQUIRE $\alpha$ (minimum duplicate factor), $\beta$ (overlap ratio), \\
  a group $G$ of $n$ accounts $a_1, \ldots, a_n$, \\
  sets $T(a_1), \ldots, T(a_n)$ of tweets where  $T(a_i) = \{t_{i1},\ldots,t_{i200}\}$ of the 200 most recent tweets for each account $a_i, 1\leq i \leq n$
 %\ENSURE  out
  \STATE $C = \emptyset$\LineComment{most frequent tweet set}
  \STATE $S = \emptyset$\LineComment{bot account set}
  \FOR {each user $a_i \in G$}
  \IF {($|\{i \mid  t \in T(a_i); 1\leq i \leq n\}| \geq \alpha$)}
  \STATE $C = C \cup \{t\}$
  \ENDIF
  \ENDFOR  
  
  \FOR {each user $a_i \in G$}
  \IF {($a_i \in S \iff \frac{|T(a_i) \cap C|}{|T(a_i)|} \geq \beta$)}
  \STATE $S = S \cup \{a_i\}$
  \ENDIF
  \ENDFOR
    
 \RETURN $C, S$
 \end{algorithmic} 
 \end{algorithm}

There are two user-specified parameters in our detection protocol: $\alpha$, which we call \textit{minimum duplicate factor}, which influences the construction of the most frequent tweet set $C$, and $\beta$, which we call the \textit{overlap ratio}, which determines the ratio of frequent tweets in the tweet set associated with an account. We performed parameter tuning and chose $\alpha=3$ and $\beta=0.6$ for our deployed detection algorithm. The results of parameter tuning are included in Appendix \ref{section:parameter_tuning}.

% Use mini figure to save space
%\begin{figure}[!h]
%\centerline{\includegraphics[width =0.5 \textwidth]{image/parameter_sweeping_plot_all_URL_shorteners.png}}
%\caption{Sensitivity of parameter overlap ratio, with minimum duplicate factor=3}
%\label{figure:overlap_ratio}
%\end{figure}

\subsubsection{URL resolver}
Once our system identifies a botnet, it extracts all embedded URLs from each account's past 200 tweets, finds the most frequent embedded URL in them, and uses Selenium\footnote{Official documentation: \url{http://www.seleniumhq.org/}} to simulate a web browser and checks if a URL is malicious or not. The detailed algorithm is in Appendix \ref{section:url_resolving_algorithm}.

We look for two types of malicious behaviors: phishing and use of secret URLs. Phishing  redirects on the client side by injecting Javascript code into a web page so that a user who clicks on URL $A$ will end up landing on URL $B$, usually containing adware or malware. In a secret URL, typing the network location part of the URL redirects a user to a safe site, but the complete URL redirects to a spam site.  For example, in our dataset we find a spam website \textit{\url{vidisp.review}}. If a user  visits \textit{\url{vidisp.review}} directly, he will be redirected to \textit{\url{google.com}}. However, if a user visits a URL containing \textit{\url{vidisp.review}} embedded in a tweet, such as \textit{\url{vidisp.review/client/bqY8G/e57Nx/avg30/1VvQj}}, he will be redirected to a spam web page.

%By our definition, malicious websites are those that cannot be resolved by programs not supporting client side Javascript execution. A typical website redirection is handled by the server returning an \textit{HTTP 301 Moved Permanently} redirect status response code, together with a new URL a program should redirect to. 

\subsubsection{Spam Campaign Classifier}
After identifying botnets and extracting malicious URLs, the final step is to identify spam campaigns. Our system queries the WHOIS database to get registrant email for every malicious URL, then it builds a dictionary that maps every registrant email address to a list of botnets whose tweets contain URLs registered under that email address. This reverse engineered map enables us to identify malicious registrants who create thousands of domains and control thousands of bot accounts. We record suspicious registrant information in a blacklist that is used by our system in subsequent detection efforts.
%and plan to share this blacklist with research community, %social media companies, and domain registrars.

\subsection{Frontend Subsystem}
%\begin{figure}[!ht]
%\centerline{\includegraphics[width = 0.5 \textwidth]{image/website_table.png}}
%\caption{A group of accounts who pretend to be Trump supporters, but who are trying to inject malware by sending shortened URLs and asking people to click URLs. Visualized by our Twitter Bot Monitor.}
%\label{figure:website_table}
%\end{figure}

Fig. \ref{figure:architecture} shows a screenshot of our web application that displays both group level and account level statistics of botnets. The group level table displays the total number of accounts in each bot group and the most frequent embedded URL tweeted by that group. The account level table includes id, screen name, statuses count, friends count, followers count, user registration language and account creation time of every account in a bot group. Along with the web portal, we also provide a complete list of Restful APIs so researchers can get archived bot statistics in JSON format by sending \textit{http} requests to our server\footnote{For our complete API Documentation, visit http://water.clear.rice.edu:18000/api.html}.

\section{Experimental Results} \label{section:Experimental_Results}
In this section we present the landscape of URL-tweeting Twitter bots and spam campaigns. We also compare our approach with existing bot detection methods. 

\subsection{Level 1: Identify botnets}
To establish a benchmark, we chose the nine most widely used Twitter URL shortening services, collected 50,000 tweets from each service, and ran our bot detection protocol. The nine URL shortening services are {\em bit.ly}, {\em ift.tt}, {\em ow.ly}, {\em goo.gl}, {\em tinyurl.com}, {\em dlvr.it}, {\em dld.bz}, {\em viid.me} and {\em ln.is}. Table \ref{table:spam_result_table} shows percentages of bot accounts identified by our system and account statuses we obtained by revisiting the Twitter API. 

\begin{table}[!ht]
\centering
\caption{Statistics of Twitter accounts from nine URL shortening services. Note the uptick in suspended accounts identified by us on {\tt viid.me}}
\begin{tabular}{l|c|c|c|c|c|c} \toprule
\pbox{20cm}{URL \\shortener}&\pbox{20cm}{Total \\\# of\\ accts}&\pbox{20cm}{Total \\\# of\\ bots}&\pbox{20cm}{\% bots \\ we\\ found}&\pbox{20cm}{\% our bots \\ susp. by \\ Twitter\\by\\ 6/10/17}&\pbox{20cm}{\% our bots \\ susp. by \\ Twitter\\by\\ 7/17/17}&\pbox{20cm}{\% our bots \\ susp. by \\ Twitter\\by\\ 9/25/17}\\ \toprule
bit.ly & 28964 & 696 & 2.40\%  & 3.74\%& 4.74\% & 8.91\%\\ \toprule
ift.tt & 12543 & 321 & 2.56\%  & 2.80\%& 9.97\% &10.59\%\\ \toprule
ow.ly & 28416 & 894 & 3.15\%   & 45.30\%& 48.21\% &48.43\%\\ \toprule
tinyurl.com & 20005 & 705 & 3.52\% & 5.39\%& 7.66\% &12.34\%\\ \toprule
dld.bz & 6893 & 304 & 4.41\%  & 8.22\%& 11.84\% &18.75\%\\ \toprule
viid.me & 2605 & 129 & 4.95\% & 38.76\%& 55.81\% &63.57\%\\ \toprule
goo.gl & 11250 & 710 & 6.31\% & 0.42\%& 3.24\% &7.04\%\\ \toprule
dlvr.it & 15122 & 1194 & 7.90\% & 7.37\%& 9.13\% &9.46\%\\ \toprule
ln.is & 25384 & 5857 & 23.07\% & 1.11\%& 1.25\% &1.50\%\\ \toprule
\end{tabular}
\label{table:spam_result_table}
\end{table}

We compared our results with BotOrNot, a supervised, account-based Twitter bot classifier. BotOrNot assigns a score of 0 to 1 to an account based on more than 1000 features, including temporal, sentiment-oriented and social network information \cite{Davis:BotOrNot}. A score closer to 0 suggests a human account, while a score closer to 1 suggests a bot account. Based on our previous work \cite{Chen:HuntingMaliciousBots}, the average BotOrNot score for the bot groups we detect ranges from 0.44 to 0.71, and in 5 out of 9 datasets, more than 50\% of BotOrNot scores fall in the range of 0.4 and 0.6, which is an indeterminate classification. Few accounts receive a score near 0 or 1. %This ambiguity comes from the nature of the probabilistic inference model that BotOrNot adopts.

%Fig. \ref{figure:BotOrNot_histogram} shows a histogram distribution of BotOrNot scores in each dataset. 

In contrast to BotOrNot, one of the biggest advantages of our unsupervised bot detection system is the ability to find new bot groups in real-time. Our system currently executes daily jobs that collect 30,000 tweets from the seven most active URL shortening services\footnote{The seven URL shortening services are {\em bit.ly}, {\em ift.tt}, {\em ow.ly}, {\em goo.gl}, {\em tinyurl.com}, {\em dlvr.it} and {\em dld.bz}. Twitter has suspended {\em viid.me} and {\em ln.is} since July, 2017.}. Fig. \ref{figure:architecture} shows daily percentages of tweets created by bots from the seven URL shortening services on Twitter over a two-month period. Tweets from bots consistently account for 10\% to 50\% of tweet traffic with shortened URLs.

%\begin{figure}[!ht]
%\centerline{\includegraphics[width = 0.48 \textwidth]{image/percent_bot_tweet_line_plot.png}}
%\caption{Percentages of daily tweets created by bots. The most abused URL shortening services is {\em dld.bz}. One botnet using {\em dld.bz} stopped to tweet for several days in the middle of September, explaining the sudden drop.}\label{figure:percent_bot_tweet}
%\end{figure}

%\begin{figure}[!ht]
%\begin{minipage}[t]{0.48 \linewidth}
%    \includegraphics[width=\linewidth]{image/percent_bot.png}
%    \caption{Percentages of bots over a month}
%\label{figure:percent_bot}
%\end{minipage}%
%    \hfill%
%\begin{minipage}[t]{0.48 \linewidth}
%    \includegraphics[width=\linewidth]{image/percent_bot_tweet.png}
%	\caption{Percentages of tweets created by bots over a month}
%\label{figure:percent_bot_tweet}
%\end{minipage}
%\end{figure}

\subsection{Level 2: Identify spam campaigns} 
\begin{figure}[!ht]
\centerline{\includegraphics[width = 0.48 \textwidth]{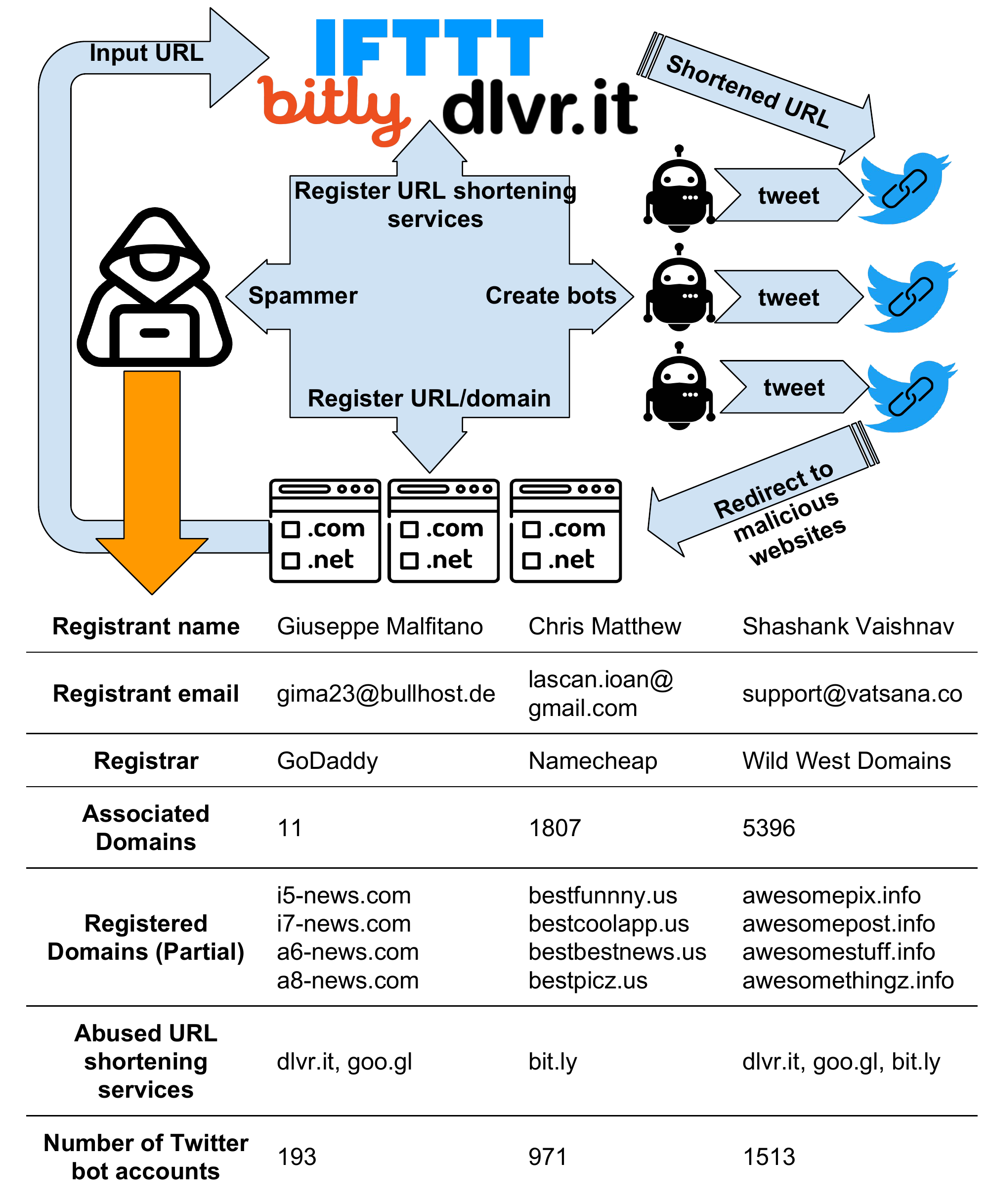}}
\caption{Top: workflow of a spam campaign. Bottom: three large scale campaigns  identified by our system. To evade detection, a spammer will create malicious websites, register for several URL shortening services, and use botnets to send shortened URLs that redirect users to these websites.} 
\label{figure:spam_campaign}
\end{figure}

We use the WHOIS database to retrieve domain registration information of every suspicious URL in every botnet. We then map registrant email to botnets to identify spam campaigns. From September 10, 2017 to November 14, 2017, our detection system  identified 848 unique suspicious registration emails behind 11,048 botnets with a total number of 185,922 accounts. %Through detailed analysis, 
We find that spammers create multiple domains as proxy websites, whose only purpose is to link to the same parent website. This link farming structure exploits search engine optimization algorithms to elevate the ranking of the parent website. A higher ranking means that more visitors will be redirected to the parent spam website, endangering more Internet users. Fig. \ref{figure:spam_campaign} visualizes the workflow of a spam campaign.

%Every domain has only one registrant, but a registrant can register multiple domains. We find registrant email to be more reliable than registrant name. The reason is that a registrant name can be a pseudonym and domain name registrars do not check its verity. The registrant email on the other hand cannot be fabricated as it is required to confirm registration. However, spammers can hide their identities by using domain proxy services, in which case the spammer's email will be replaced by a proxy email. Among all 1,351 suspicious URLs we have collected 19.32\% uses proxy. 

The scale of these spam campaigns is often  very large. Fig. \ref{figure:spam_campaign} shows statistics of three large-scale campaigns we have identified, whose recorded registration names are Giuseppe Malfitano, Chris Matthew, and Shashank Vaishnav\footnote{These names are provided by the WHOIS database. Since many domain registrars do not check the validity of registration name, these could well be fictitious.}. All three spammers create phishing websites, and one of them (Chris Matthew) has also created websites with secret URLs. They register a large number of domains that contain attractive words (such as \textit{news}, \textit{best}, \textit{awesome}) to lure users to click their links. Many domains are not live yet which means that they are registered but do not have associated IP addresses. We believe that these dormant domains will become active to serve spam content once the currently active domains are blocked. For this reason it is important for domain registration companies and social media companies to preempt spam activities by blocking both active and non-active domains registered by suspicious registrants such as these, identified by our system.

%\begin{table}[!ht]
%\centering
%\caption{Registration information of three spam campaigns and their links to URL shortening services and Twitter botnets}
%\label{table:three_spam_campaigns}
%\begin{tabular}{lrrrr}
%\toprule
%\pbox{20cm}{Spam website \\owner}
%& Giuseppe Malfitano
%& Chris Matthew                                                                                
%& Shashank Vaishnav \\ \toprule
%\pbox{20cm}{Owner\\ email\\address}                       & gima23@bullhost.de
%& \pbox{20cm}{lascan.ioan@\\gmail.com}
%& support@vatsana.co
%\\ \toprule
%\pbox{20cm}{Registrar}
%& GoDaddy
%& Namecheap 
%& Wild West Domains
%\\ \toprule
%\pbox{20cm}{Associated\\ Domains} 
%&11
%&1807
%&5396
%\\ \toprule
%\begin{tabular}[c]{@{}c@{}}Domains\\registered\\(partial)\end{tabular}& \begin{tabular}[c]{@{}l@{}}i5-news.com\\i7-news.com\\a6-news.com\\a8-news.com\end{tabular} & 
%\begin{tabular}[c]{@{}l@{}}bestfunnny.us\\bestcoolapp.us\\bestbestnews.us\\bestpicz.us	\end{tabular} & 
%\begin{tabular}[c]{@{}l@{}}awesomepix.info\\awesomepost.info\\awesomestuff.info\\awesomethingz.info\end{tabular}  \\ \toprule
%\pbox{20cm}{URL\\ shortening\\services\\abused}                                 & dlvr.it, goo.gl
%& bit.ly
%& dlvr.it, goo.gl, bit.ly \\ \toprule
%\pbox{20cm}{Number of\\ Twitter bots}                                 & 193
%& 971 
%& 1513 \\ \toprule
%\end{tabular}
%\end{table}

\section{How do spammers control botnets? Case study: from creating bot accounts to hijacking normal accounts}
\label{section:bot_mechanism}
In this section, we describe two general mechanisms that spammers use to control botnets. A traditional approach is to register a large number of accounts under a single user name, and a more unconventional approach is to build a Twitter app to hijack normal users' credentials. Current bot detection methods that operate at the level of an individual account  have difficulty detecting users comprised by the second mechanism.  In contrast, our system  correctly identifies coordinated bot activities resulting from both mechanisms weeks before Twitter detects and flags or suspends these accounts.  Here we document four case studies that show how those two bot control mechanisms work on Twitter.

%A traditional bot group consists of a large number of accounts registered by a miscreant and programmed to tweet the same content or URL so the entire group is homogeneous. To remove this kind of botnet a supervised model with good features or an unsupervised model with right underlying assumptions will be sufficient. More recently, a spammer will first create a Twitter application (usually called Twitter app), then  persuades legitimate Twitter users to authorize the app to tweet on their behalf. Once this application is authorized, it can launch large scale bot activity on Twitter. In this case detecting individual account is inefficient because most are innocent accounts hijacked by the application. Our unsupervised system has the ability to detect this type of bot activity in real-time. Since its deployment on February 2017, our system has successfully identified onsets of coordinated bot activities weeks before Twitter took any anti-spam action. Here we document four case studies that show how coordinated bot activities evolve on Twitter.

\subsection{Traditional bot groups (savingzev and club)}
On February 25, 2017, our system detected a large number of trending URLs that use {\em .us} as their top-level domain, a variety of second-level domains, and the same subdomain {\em savingzev} (for example: {\em savingzev.feedsted.us}). These URLs all redirected to a phishing website with low quality content mixed with links to malware. Two weeks later, on March 27, 2017, Twitter introduced a link unsafe warning message before redirecting the user to those sites. 

On April 13, 2017, our system detected a list of URLs with {\em .club} as top-level domain and a range of second-level domains (for example: {\em likelisi.club}). Similar to {\em savingzev}, the final landing URL pointed to a phishing news website. 
%The tweet volume went down after several days (*** SPECIFY EXACTLY WHEN and WHY ***) and 
When we revisited these URLs in September 2017 they were no longer valid domain names and Twitter appeared to have suspended 83\% of the associated accounts.

\renewcommand{\arraystretch}{1.2}
\begin{table}[!ht]
\centering
\caption{Domain registration information of four Twitter bot groups}
\label{table:domain_registration_four_groups}
\begin{tabular}{cllll}
\midrule
\textbf{Bot group} &du3a & twitbr &savingzev  & club \\     
\midrule
\pbox{20cm}{\relax\ifvmode\centering\fi\bf Date\\ detected}
&Long-lasting 
&09/28/17 
&02/25/17  
&04/13/17
\\   
\midrule
\begin{tabular}[c]{@{}c@{}}\pbox{20cm}{\relax\ifvmode\centering\fi\bf Domains\\detected\\(partial)}\end{tabular} & \begin{tabular}[c]{@{}l@{}}du3a.org\\d3waapp.org\\ghared.com\end{tabular} & \begin{tabular}[c]{@{}l@{}}twits.tk\\twitbr.tk\\viraltt.tk\end{tabular} & \begin{tabular}[c]{@{}l@{}}savingzev.renewsfeed.us\\savingzev.qualifystory.us\\savingzev.feedsted.us\end{tabular} & \begin{tabular}[c]{@{}l@{}}takensi.club\\ takenst.club\\ likelino.club\\ likelisi.club\end{tabular} \\ \toprule
\pbox{20cm}{\relax\ifvmode\centering\fi\bf Top level \\domain}                                & \begin{tabular}[c]{@{}c@{}}.org\\ .com\end{tabular}                                          & .tk                                                                       & .us                                                                                                       & .club
\\
\midrule
\pbox{20cm}{\relax\ifvmode\centering\fi\bf Total \# \\ of accts}                     
& 38262
& 4706                                                               & 1243                                                                                                   & 3229 
\\ \midrule
\pbox{20cm}{\relax\ifvmode\centering\fi\bf Total \# \\ of bots}                       
& 32591                                                                                 & 1526                                                               & 1238                                                                                                   & 941                                                                                              \\ \midrule
\pbox{20cm}{\relax\ifvmode\centering\fi\bf \% bots}                       
& 85.18\%
& 32.34\%                                                           & 99.60\%                                                                                                & 29.14\%                                                                                                 \\ \midrule
\pbox{20cm}{\relax\ifvmode\centering\fi\bf \% tweets\\ from bots}     
& 85.03\%
& 29.82\%                                                           & 99.99\%                                                                                                & 85.70\%                                                                                                 \\ \midrule
\pbox{20cm}{\relax\ifvmode\centering\fi\bf \% our bots\\ susp. by \\Twitter\\ until 10/01/17}                       & 00.00\%                                                                                 & 00.00\%                                                               & 96.28\%                                                                                             & 83.21\%                                                                                             \\ \midrule
\end{tabular}
\end{table}

Traditional bot groups register domains and Twitter accounts in batches \cite{Hao:PREDATORProactiveRecognition}. Table \ref{table:domain_registration_four_groups} shows that domain names of \textit{savingzev} and \textit{club} groups have similar lexical features. \textit{Savingzev} abuses subdomain registration by adding subdomain \textit{savingzev} before every second-level domain as a unique address without having to purchase a new domain name. Similarly, \textit{Club} group abuses top-level domain providers by registering with a lesser-known domain {\em .club}. Previous research \cite{Thomas:SuspendedAccountsinRetrospect} also notes that spammers ``rely on free subdomains to avoid registration costs'', or use subdomains to inflate search engine results on certain keywords. We looked up domain registration information on WHOIS database for all top 20 trending domains on Twitter containing subdomain \textit{savingzev} on February 25, 2017 and found they were all created by one registrant with one email address on February 24, 2017, one day before the launch of Twitter bot activity.

%\textit{Club} group abuses top-level domain providers by registering with a lesser-known domain {\em .club}. According to 2017 IBM Threat Intelligence Report \cite{IBM:ThreatIntelligence}, these new and low-cost domains allow spammers to bypass URL filters and register hundreds of domains in a short time.

\subsection{Unconventional bot groups (du3a and twitbr)}
A new mechanism for spreading spam and malware on Twitter is that a malicious Twitter user will first create a Twitter app, and then encourage ordinary Twitter users to subscribe to the service provided by the app. After getting authorized, the app  reads tweets from the user's timeline, gets the user's followers, follows new accounts and post tweets on the user's behalf. With enough number of subscribed users, a malicious user has the power and resources to launch bot activity by hijacking and abusing subscribed accounts, until a user rechecks his or her account and disables the app.

On September 28, 2017, our real-time Twitter abnormal trend monitor detected {\em twitbr.tk}, a social media managing app,  entering the top $k$ URL list. This triggered our bot detection protocol with keyword {\em twitbr}. We observed  Twitter users complaining about {\em twitbr.tk} on social media and posting solutions on how to disable the app. To understand how this app spreads spam, we subscribed to the app using an experimental Twitter account. Seconds after the registration, the app started to tweet malicious links, and  begin tagging friends of our experimental account. Three URLs, {\em twits.tk}, {\em twitbr.tk}, and {\em viraltt.tk} were used to launch this spam activity. The uncommon top-level domain {\em .tk} is abused in a similar manner as {\em .club}. As of October 22, 2017, Twitter  flagged {\em twitbr.tk} as dangerous and has taken down the application, but other variations of such URLs are still trending on Twitter.

Also belonging to this group is a family of prayer tweeting apps, whose domain names include {\em du3a.org, d3waapp.org, ghared.com, zad-muslim.com}. Those apps tweet in Arabic, have a large user base, and are long standing members of our top $k$ URL list.  Legitimate Twitter users authorize these apps to send prayers on their behalf. Although the prayer tweeting service is innocuous and Twitter has permitted them to function unimpeded, a recent study by Berger and Morgan stated that ``some ISIS supporting Twitter accounts used bots and apps to facilitate their activity on Twitter.\cite{Berger:TheISISTwitterCensus}''  Berger and Morgan also noted that ``these apps introduce noise into social networks and their use may be intended to impede analysis.''

%The user can also customizes prayers and locations [refre to user agreement of http://www.athantweets.com/].

\subsection{Results and Comparisons}

\begin{figure}[!ht]
\begin{minipage}[t]{0.5\linewidth}
    \includegraphics[width=\linewidth]{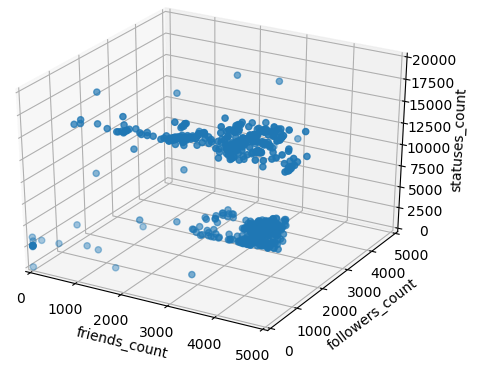}
\end{minipage}%
    \hfill%
\begin{minipage}[t]{0.5\linewidth}
    \includegraphics[width=\linewidth]{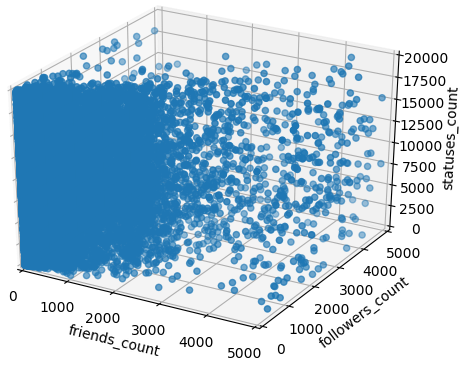}
\end{minipage} 
\begin{minipage}[t]{0.5\linewidth}
    \includegraphics[width=\linewidth]{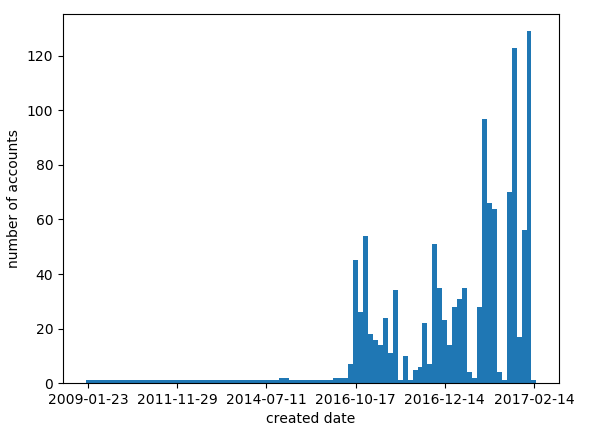}
    %\caption{...add}
    %\label{figure:plot_3d_twits}
\end{minipage}%
    \hfill%
\begin{minipage}[t]{0.5\linewidth}
    \includegraphics[width=\linewidth]{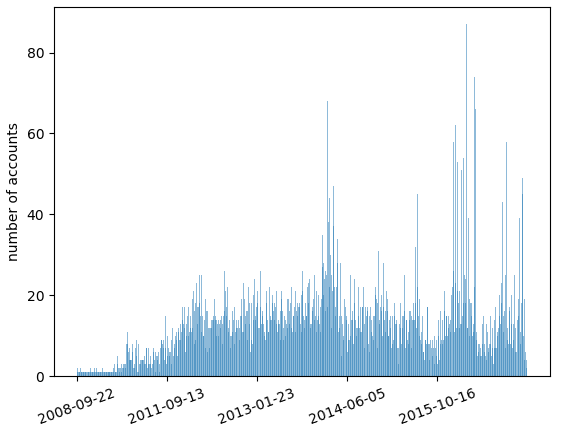}
    %\caption{... add}
    %\label{figure:plot_3d_du3a}
\end{minipage} 
	\caption{Scatter plots (statuses count, friends count and followers count) and account creation time plots of \textit{savingzev} (left column) and \textit{du3a} (right column) groups. See Sections V.A. and V.B. for explanation.}
	\label{figure:plot_3d_twits_du3a}
\end{figure}

% \begin{figure}[!ht]
% \begin{minipage}[t]{0.5\linewidth}
%     \includegraphics[width=\linewidth]{image/created_at_bar_chart_by_day_savingzev.png}
%     %\caption{...add}
%     %\label{figure:created_at_bar_chart_by_day_savingzev}
% \end{minipage}%
%     \hfill%
% \begin{minipage}[t]{0.5\linewidth}
%     \includegraphics[width=\linewidth]{image/created_at_bar_chart_by_day_club.png}
%     %\caption{... add}
%     %\label{figure:created_at_bar_chart_by_day_club}
% \end{minipage} 
% 	%\caption{Account creation time (left: savingzev, right: club)}
% 	%\label{figure:created_at_savingzev_club}
% %\end{figure}
% %\begin{figure}[!ht]
% \begin{minipage}[t]{0.5\linewidth}
%     \includegraphics[width=\linewidth]{image/created_at_bar_chart_by_day_twits.png}
%     %\caption{...add}
%     %\label{figure:created_at_bar_chart_by_day_twits}
% \end{minipage}%
%     \hfill%
% \begin{minipage}[t]{0.5\linewidth}
%     \includegraphics[width=\linewidth]{image/created_at_bar_chart_by_day_du3a.png}
%     %\caption{... add}
%     %\label{figure:created_at_bar_chart_by_day_du3a}
% \end{minipage} 
% 	\caption{Account creation time (Top left to bottom right: savingzev, club, twitbr, du3a). Notice the difference in the y-axis scale: accounts from top two groups are created in bulk.}
%     \label{figure:created_at_twitbr_du3a}
% \end{figure}

We now detail how Twitter has responded to these four groups and explain differences between traditional botnets and botnets assembled from hijacked accounts. Table \ref{table:domain_registration_four_groups} shows the number of bot accounts identified by our unsupervised approach and their statuses after revisiting Twitter API. Note that even though all four groups have high proportions of bot accounts, Twitter aggressively suspended \textit{savingzev} and \textit{club} groups but did not suspend accounts from \textit{du3a} and \textit{twits} groups, because these accounts are owned by legitimate users who are unlikely to be aware of their accounts' activities.

To demonstrate that traditional bot groups are much more homogeneous than unconventional groups, we collected account profile data from Twitter Rest API. Fig. \ref{figure:plot_3d_twits_du3a} shows the scatter plot of statuses count, friends count and followers count and account creation time. It is clear that \textit{savingzev} accounts are highly clustered and are created in a batch by one or a few people, while \textit{du3a} groups are randomly distributed in the 3D space and have a more uniform creation time distribution.

Even though these two bot coordination mechanisms are very different, our group-based, unsupervised approach is able to detect spam activity generated by both of them in real-time. Further, we identify the root cause of the threat, i.e., whether it is from a Twitter application hijacking user accounts or a  botnet created by conventional means.

\section{Conclusion} \label{section:conclusion}
We design and implement an unsupervised  system to detect Twitter spam campaigns that use bot groups to send duplicate content with embedded URLs. We show that bots consistently account for 10\% - 50\% of tweet traffic from URL shortening services by conducting measurements using the Twitter Streaming API over a two-month period. By mapping domain registration information to botnets, we find large scale spam campaigns that control thousands of malicious websites and bot accounts. We illustrate two different mechanisms spammers use to control Twitter botnet: creating a large number of accounts registered under a single user, and creating a single Twitter application that co-opts legitimate accounts. Finally, we create a web application that monitors the top $k$ trending URLs on Twitter and visualizes the botnets and spam campaigns we have identified. We  provide API interfaces for easy access
to the database of such botnets. We hope that our findings can reduce spam and malware on Twitter and  improve the quality of Twitter datasets used for social network and content analysis.

% or


\appendices
\section{Parameter Tuning} \label{section:parameter_tuning}
Minimum duplicate factor ($\alpha$) and overlap ratio ($\beta$) are two parameters used in our detection protocol. High values of $\alpha$ and $\beta$ correspond to a more homogeneous bot group. If we set those two parameters too low, the false positive rate will increase, whereas if we set them too high, the false negative rate will increase. To find an appropriate setting, we apply elbow method where we choose a number such that increasing the value does not affect much about the outcome.

\begin{figure}[!ht]
\begin{minipage}[t]{0.48 \linewidth}
    \includegraphics[width=\linewidth]{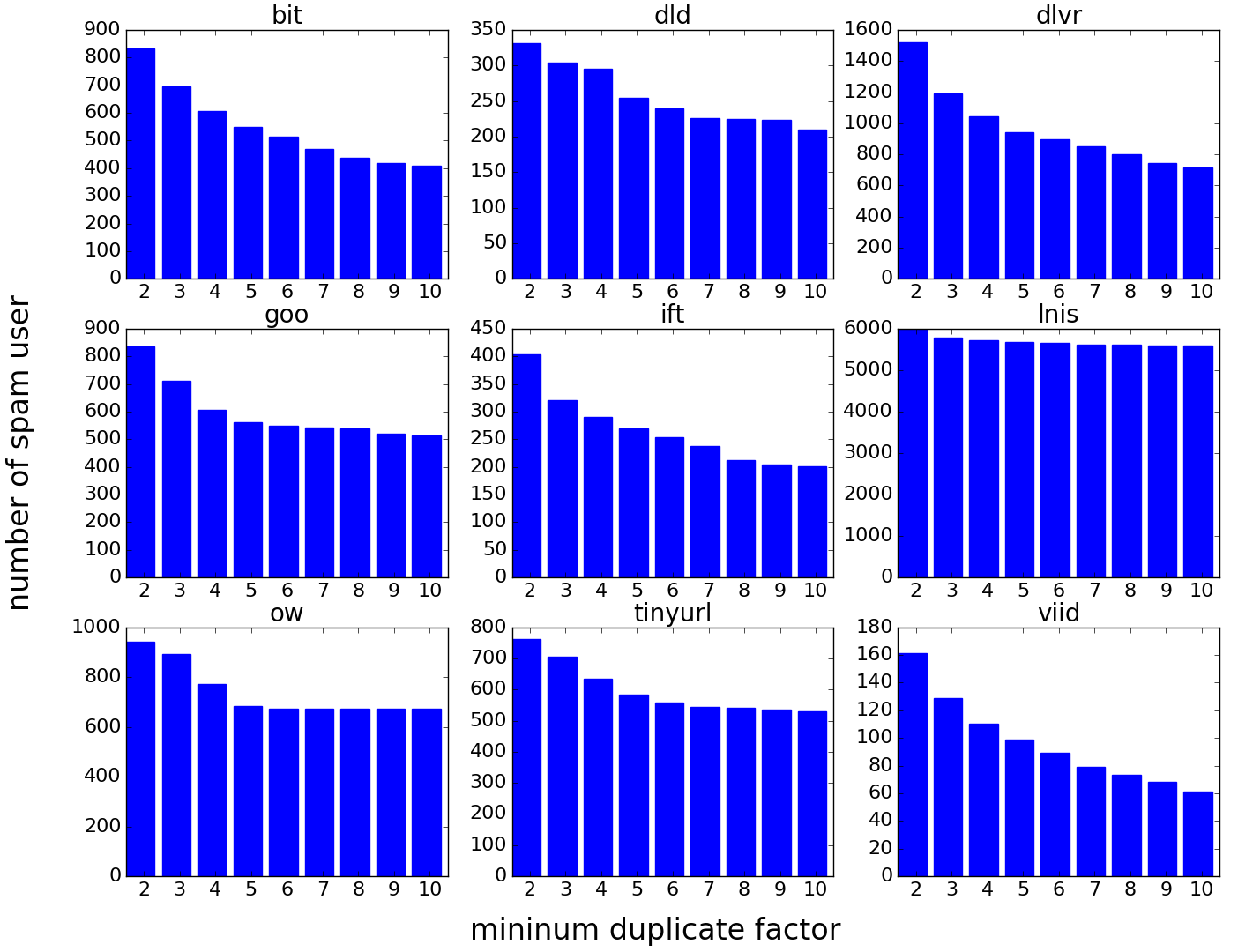}
    \caption{Variation in number of detected bots as a function of  minimum duplicate factor with fixed overlap ratio = 0.6}
\label{figure:minimum_duplicate_factor}
\end{minipage}%
    \hfill%
\begin{minipage}[t]{0.48 \linewidth}
    \includegraphics[width=\linewidth]{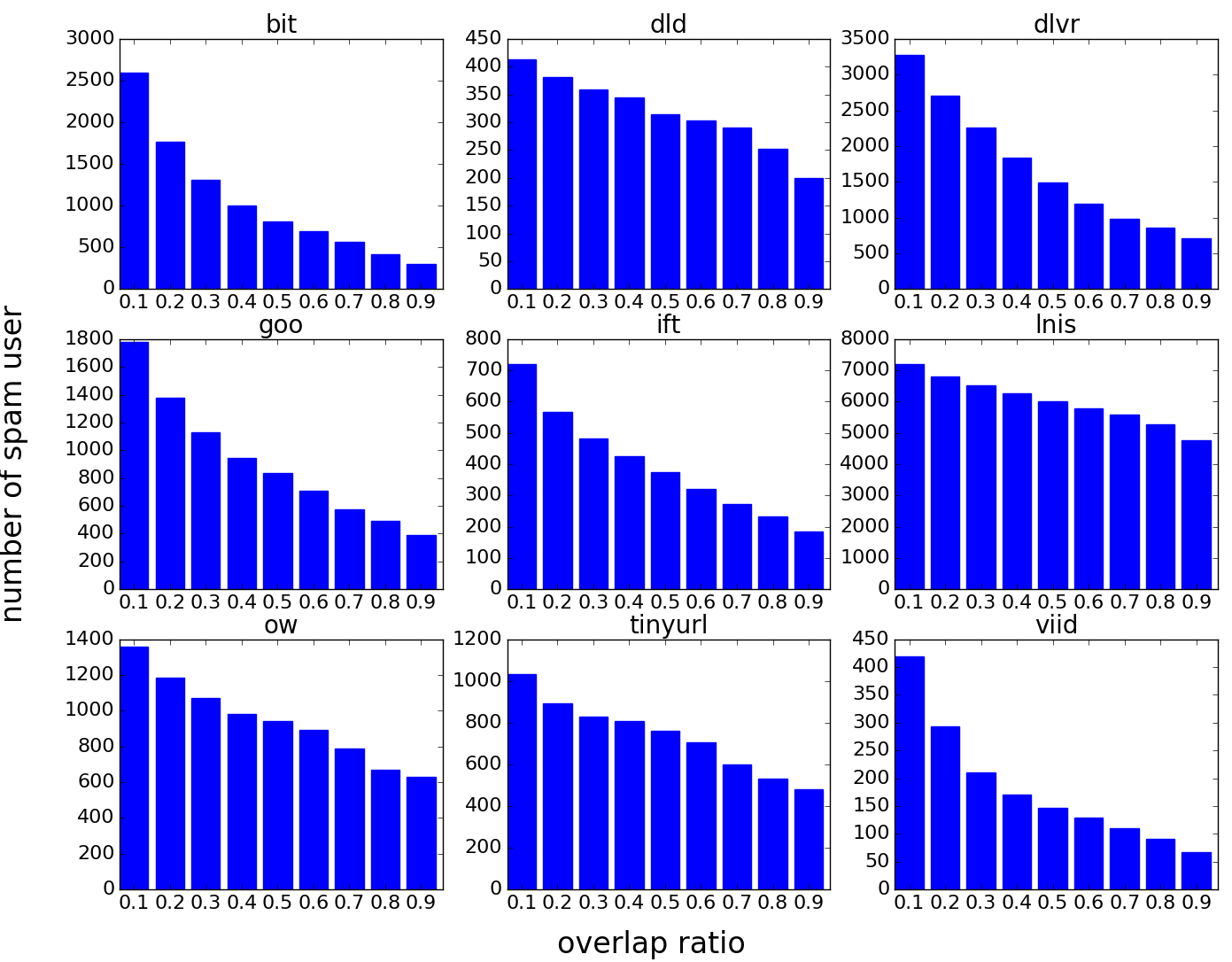}
	\caption{Variation in number of detected bots as a function of  overlap ratio with fixed minimum duplicate factor = 3}
\label{figure:overlap_ratio}
\end{minipage}
\end{figure}

%\subsubsection{Minimum Duplicate Factor $\alpha$}
Fig. \ref{figure:minimum_duplicate_factor} shows the sensitivity of minimum duplicate factor $\alpha$, with a fixed overlap ratio of 0.6. Other than {\em viid.me} dataset, performance of the detection method is not sensitive to the choice of minimum duplicate factor. The largest marginal decrease of bot accounts is from 2 to 3, so we choose an $\alpha=3$ as the minimum duplicate factor.

% Use mini figure (two columns to save space)
%\begin{figure}[!h]
%\centerline{\includegraphics[width =0.5 \textwidth]{image/parameter_sweeping_plot_all_URL_shorteners_min_dup_factor.png}}
%\caption{Sensitivity of parameter minimum duplicate factor, with overlap ratio=0.6}
%\label{figure:minimum_duplicate_factor}
%\end{figure}

%\subsubsection{Overlap Ratio $\beta$}
Fig. \ref{figure:overlap_ratio} shows the sensitivity of overlap ratio, with a fixed minimum duplicate factor equal to 3. Interestingly, this parameter is more sensitive to URL shortening services {\em bit.ly}, {\em dlvr.it}, {\em goo.gl}, {\em ift.tt} and {\em viid.me}, but less sensitive to {\em dld.bz}, {\em ln.is}, {\em ow.ly} and {\em tinyurl.com}. This suggests that accounts using the first five URL shortening services are more heterogeneous with fewer identical tweets, while accounts tweeting from the other four services are more homogeneous. An exception is {\em viid.me} dataset, whose bot accounts are all injecting malwares, but do not always tweet duplicate content. This group is more sophisticated and requires further investigation. For all URL shortening services, the number of bot accounts decreases as the overlap ratio increases. We choose an overlap ratio of 0.6 because that is where plots start to plateau.

\section{URL Resolving Algorithm} \label{section:url_resolving_algorithm}

 \begin{algorithm}
 \caption{Algorithm for resolving final landing URL}
 \label{alg:url_resolve}
 \begin{algorithmic}[1]
 \renewcommand{\algorithmicrequire}{\textbf{Input:}}
 \REQUIRE a URL $topURL$
 %\ENSURE  out
  \STATE $numRetry$ = 0, $maxRetry$ = 5, $code$ = 301
  \STATE $has\_secret\_url$ = false, $is\_phishing$ = false
  \WHILE{$code$ == 301 and $numRetry < maxRetry$}
  \STATE response = request.get($topURL$)
  \STATE $code$ = response.status\_code
  \STATE $topURL$ = response.url, 
  \STATE $numRetry$ += 1
  \ENDWHILE \\
  \LineCommentLeftAlign{nlp means network location part}
  \STATE $u1$ = selenium($topURL.nlp$)
  \STATE $u2$ = selenium($topURL$)
  \IF {($u1.nlp \ne u2.nlp$)}
  \STATE $has\_secret\_url$ = true
  \ENDIF
  \IF {($u1.nlp \ne topURL.nlp $ or $u2.nlp \ne topURL.nlp $)}
  \STATE $is\_phishing$ = true
  \ENDIF

 \RETURN $topURL$, $has\_secret\_url$, $is\_phishing$ 
 \end{algorithmic} 
 \end{algorithm}

%%%%%%%%%%%%%% UNUSED -- bot or not score histogram %%%%%%%%%%%%%%
% \begin{figure}[!ht]
% \centerline{\includegraphics[width = 0.5 \textwidth]{image/bot_or_not_api_score_all_URL_shorteners.png}}
% \caption{Histogram of BotOrNot score for nine datasets. Most scores fall in the range of 0.4 to 0.6, a range of uncertainty.} 
% \label{figure:BotOrNot_histogram}
% \end{figure}
%%%%%%%%%%%%%%%%%%%%%%%%%%%%%%%%%%%%%%%%%%%%%%%%%%%%%%%%%%%%%%%%%%

%\section*{Acknowledgment}
%The authors would like to thank

\ifCLASSOPTIONcaptionsoff
  \newpage
\fi



%\nocite{*}
\bibliographystyle{IEEEtran}

\bibliography{main.bib}


%\begin{biography}[{\includegraphics[width=1in,height=1.25in,clip,keepaspectratio]{mshell}}]{Michael Shell}

%\begin{IEEEbiography}[{\includegraphics[width=1in,height=1.25in,clip,keepaspectratio]{picture}}]{Zhouhan Chen}
%is a student from Rice University
%\end{IEEEbiography}


%\vfill


\end{document}